\documentclass[11pt]{article}
\usepackage{amssymb,amsmath,latexsym,amscd,amsfonts}  
\usepackage{graphics}  
\usepackage{epsfig}  
  
\newtheorem{theorem}{Theorem}[section]  
  
\newtheorem{definition}[theorem]{Definition}

\newtheorem{lemma}[theorem]{Lemma}

\newcommand{\qedsymb}{\hfill{\rule{2mm}{2mm}}}  
\newenvironment{proof}[1][]{\begin{trivlist}  
\item[\hspace{\labelsep}{\bf\noindent Proof#1:\/}] 
}{\qedsymb\end{trivlist}}  
  
\def\C{\mathbb{C}}

\def\ra{\rangle}  
\def\la{\langle}

\newcommand{\znote}[1]{{\bf (Zeph:} {#1}{\bf ) }}  
  
\newcommand{\ignore}[1]{}  
  
\renewcommand{\epsilon}{\varepsilon}

\def\hpic #1 #2 {\mbox{$\begin{array}[c]{l} 
\epsfig{file=#1,height=#2} \end{array}$}}  
\def\vpic #1 #2 {\mbox{$\begin{array}[c]{l} \epsfig{file=#1,width=#2} 
\end{array}$}}

\def\C{\mathbb{C}}

\begin{document} 
\title{The quantum FFT can be classically simulated\footnote{
This result was presented in informal discussions in QIP 2006, Paris.} 
\author{  
 Dorit Aharonov  
 \thanks{School of Computer Science and Engineering,  
 The Hebrew University, Jerusalem, Israel.  
 doria@cs.huji.ac.il. Research supported by ISF grant 032-9738 and 
Alon Fellowship}  
 \\and  
 Zeph Landau \thanks{Department of Mathematics, City College, NY} 
\\and Johann Makowsky \thanks{Department of Mathematics, The Technion,  
Haifa}}}
\maketitle 
 
 
\begin{abstract} 
In this note we describe a simple and intriguing observation: 
the quantum Fourier transform (QFT) over $Z_q$, 
which is considered the most 
``quantum'' part of Shor's algorithm, 
 can in fact be simulated efficiently by classical computers. 

More precisely, we observe that the QFT
can be performed by a circuit of poly-logarithmic path-width, if the circuit 
is allowed to apply not only unitary gates but also 
general linear gates.
Recalling the results of Markov and Shi \cite{shi} and Jozsa \cite{jo} 
which provided classical simulations of such circuits 
in time exponential in the tree-width, this implies 
the result stated in the title.  

Classical simulations of the FFT are of course meaningless when applied 
to classical input strings on which their result is already known; 
Our observation might be interesting only in the context in which the 
QFT is used as a subroutine and applied to more interesting superpositions. 
We discuss the reasons why this idea seems to fail to  
provide an efficient classical simulation of the  
entire factoring algorithm. 

In the course of proving our observation, we provide two 
alternative proofs of the 
results of \cite{shi,jo} which we use. 
One proof is very similar in spirit to that of \cite{shi} 
but is more visual, and is  
based on a graph parameter which we call the ``bubble width'', 
tightly related to the path- and tree-width. 
The other proof is based on connections to the Jones 
polynomial; It is very short, if one is willing to rely on 
several known results.  
\end{abstract}  
  
\section{Introduction}
In our attempts to understand and characterize the quantum 
computational power, it is interesting to 
understand which parts of quantum computation 
are truly {\it quantum}, and which can be simulated efficiently by classical 
computers. This has been the subject of many works over the past few years, 
e.g., the Gottesman-Knill theorem \cite{nielsen}, 
providing a simulation of quantum circuits that use only Clifford group gates, 
the simulations by Vidal of quantum circuits that use only limited amount of  
entanglement \cite{vidal},  
and the efficient simulation of circuits using 
only ``match gates'' \cite{valiant, terhal}.
  
Of particular interest in our context is the 
recent work of Markov and Shi \cite{shi} who considered
quantum circuits restricted not in the type of  
gates they use, but rather in the topology of  the graph corresponding to  
the quantum circuit (the graph whose nodes corresponds to 
 the quantum gates, and whose edges correspond 
 to the wires in the circuit).  They show that 
quantum circuits can be simulated classically in time polynomial 
in the number of gates and exponential in a topological parameter 
called the tree-width of the circuit graph. 
In \cite{shi}, Markov and Shi raised the question of whether the Quantum 
Fourier transform can be assigned small tree-width quantum circuits, 
which would imply its efficient classical simulation by their theorem.
 
In this note we observe that a simple generalization of the results 
of Markov and Shi \cite{shi} allows us to do this, namely, to show that 
 the quantum Fourier transform can be simulated classically in polynomial 
time. To state our result precisely requires a little more
 detail which we provide now.

Our approach begins by introducing a topological parameter of
 a graph called  the {\it bubble-width} of the graph.  It will 
turn out that the bubble width is closely related to the tree 
width but we find that the bubble width is a more visual parameter 
that is easier to work with.   It is defined roughly as follows:  
Imagine the graph is embedded in $R^3$ in some way, and  
that a huge spheric bubble sits very far away from the graph.  
The bubble approaches and ``eats'' the nodes of the graph one by one  
until eventualy it has swallowed the entire graph which now sits inside it.  
We think of the  
edges of the graph, and of the surface of the bubble, as flexible objects, 
made of rubber, say, and so in the process of the swallowing, both the  
surface of the bubble and the nodes and edges of the graph  
can be moved, stretched, or bent, in a continuous manner. 
In topological language, we allow isotopies of the bubble and of the graph.  
The goal is to find a way for the bubble  
to swallow the graph, such that  
the number of edges of the graph that cross the surface of the bubble at  
any given point does not exceed a certain number $c$.  
The minimal number $c$ for which such swallowing is possible is called  
the bubble-width of the graph.  

We shall consider a much more general class of  
circuits than quantum circuits which we call  
{\it operator circuits}.  
In such circuits, the gates operate on the $n$-fold tensor product of  
two dimensional vector spaces, the same space as the Hilbert space of $n$ qubits. However, the gates which we allow are not necessarily unitary  
gates, or even quantum permissable gates, i.e., completely positive maps.  
In fact, we simply allow any linear transformation from $k$ to $\ell$ qubits. 
Just like in the case of quantum circuits, there are $n$ input bits (some of 
which might be constant) and there are $m$ output bits, one of which 
is marked to be the answer of the computation. 

For an operator circuit we show the following:

\begin{theorem}\label{thm:main}  
Given an operator circuit $Q$, denote the graph associated with it by  
$G_Q$. Let $BW(G_Q)$ be the bubble-width of this graph.  Given an input string $x$, denote by $Q(x)_0$ the vector that is the projection of $Q$ applied to $x$ onto the subspace that has the answer qubit $0$.  
There exists a classical efficient algorithm that outputs the 
exact norm squared of  $Q(x)_0$; 
moreover, the time that  
the simulation takes is at most exponential in $BW(G_Q)$, and polynomial 
in the number of gates in $Q$.   
\end{theorem}  

We note that following similar arguments to the proof of Theorem 
\ref{thm:main}, we can actually also calculate 
 the exact inner product of $Qx$ with any output string $y$.

Theorem \ref{thm:main} 
 is essentially the result in \cite{shi} though that result 
is stated for tree-width instead of bubble width and 
is restricted to quantum circuits $Q$ instead of operator 
circuits (the proof of \cite{shi} works for operator circuits, a 
fact undoubtably known by the authors).  Alternate proofs of 
similar results to the above were given in \cite{jo}.  Here we 
provide yet two more proofs of the above result.  The first, 
contained in section \ref{s:3}, is self-contained and gives a 
clear picture of the association of the bubble width with the 
computation of the circuit.  The second, contained 
in section \ref{s:4}, only holds for case when $Q$ is 
indeed a quantum circuit, and uses the intriguing connection between 
quantum circuits and the Jones polynomial \cite{arad}.
This proof is very short if one is willing to rely on results 
from those areas.

We then turn to the quantum Fourier transform and show:
 \begin{theorem}\label{thm:fft}   
There exists an operator circuit which applies the quantum 
 Fourier transform on $n$ qubits to within precision $O(1/n)$  
and whose bubble width is $O(log^2(n))$.  
\end{theorem} 
The design of the operator circuit is based  
on a result of Cleve and Watrous \cite{cw} who gave  
logarithmic depth (non-planar)  
quantum circuits for the Fourier transform. Their circuits  
were of linear bubble width, but using the relaxation from  
unitary quantum circuits to operator circuits, we can show  
how to make the bubble width polylogarithmic. 

The combination of these two theorem has an 
intriguing conclusion: the quantum Fourier transform has an efficient classical simulation.  Of course, there is not much we can learn from  
applying the Fourier transform circuit on a classical input string, 
and studying the probability for some output;  
we already know that for any classical input string, the  
outcome will be distributed uniformly.   
The above statement is thus of little meaning in the context of  
classical inputs. The reason it might be of interest never the less  
is because of the hope to apply it to more interesting circuits,  
which may include the Fourier transform as a subroutine. 
The above statement shows that there is reason to believe  
that the Fourier transform part in the circuit 
will not be the obstacle towards classical  
simulations of such circuits.  
 
At this point the reader 
 might wonder why this result does not imply that factoring  
can be performed classically, since it seems that the quantum Fourier  
transform is the only truly quantum part of Shor's algorithm, i.e,  
the only part that is hard to simulate classically.  
The problem is extending the result to the entire Shor's algorithm  
lies in the first part of Shor's algorithm, namely,  
the modular exponentiation, which seems like a ``classical'' part.  
Even though the circuit is classical, it is performed on a superposition  
of all strings, and so we cannot simply simulate it by a classical  
circuit of the same size. The problem in attempting to use our methods 
 is that we would need to show how to perform the modular exponentiation 
so that the resulting circuit, and moreover, the combined circuit with the 
QFT circuit, has small bubble width.

An interesting open question is to ask whether these results can be  
used in other contexts.  
One way that one might hope to use this is in order to  
estimate the Fourier coefficients of interesting quantum states;  
if a quantum state can be generated  
with a small bubble width circuit, and if the Fourier transform  
subroutine does not increase the bubble width significantly (as is the case for instance for states coming from log-depth planar circuits),  then  
the Fourier coefficients of the state can be calculated  
efficiently classically. This might be a way to   
derive efficient classical algorithms for 
certain tasks, by first constructing  
a small bubble-width operator circuit for the task.

\bigskip
\noindent{\bf Related work}

After completing this work, we have learned that similar results
were achieved independently by Yoran and Short around the same time
\cite{yoran}. The results in \cite{yoran} are in fact somewhat stronger,
as they achieve not only quasi-polynomial simulation of the
QFT but rather a polynomial simulation. The methods are different, and we
believe there is independent merit for both results.

\section{Graph Parameters: Bubble Width, Tree Width, Path Width} 
\subsection{Notation} 
For a finite set $S$, $|S|$ will denote the number of elements of $S$. 
Given  a finite graph $G$, we shall denote by $v(G)$ the vertices of $G$ and $E(G)$ the edges of $G$.  For a given graph $G$ and vertex $v$, the star graph $G_v$ shall be the subgraph of $G$ consisting only of the edges and vertices in $G$ connected to $v$. 
 
\subsection{Bubble Width} 
\begin{definition} {\bf Bubble Width} Given a graph, a bubbling  $B$ of $G$ shall mean an  
ordering of all the vertices of $G$,  
\[ b_1, b_2, \dots ,b_n. \] 
This ordering induces a sequence of subsets  
\[S_1\subset S_2 \subset\dots \subset S_n\] 
with $S_i=\{b_1,\ldots, b_j\}$.  
For each $i$, we define $z_i(B)\subset E(G)$ to be the set of edges  
with exactly one endpoint in $S_i$.  The {\em width } of $B$  
shall be $\max_{i} |z_i(B)|$. The bubble  
width of $G$, denoted $BW(G)$, is defined  
to be the minimal width over all bubblings of $G$.  
\end{definition} 
 
\subsection{Tree-width, path width and the connection to Bubble-width} 
We show that the parameter bubble-width is tightly related to the
 well studied notions \cite{bod} of tree-width and path width. 
 
\begin{definition}  
{\bf Tree-Width, Path Width}  
A tree decomposition of a graph $G$ 
 is an undirected tree $T$, where each node $t\in T$  
is assigned a subset $\tilde{t}$ of the nodes of $G$.  
The condition for this to be a tree-decomposition is 

\begin{enumerate} 
\item For each edge $(v,w)$ in $G$, there must exist a node $t \in T$  
whose subset contains both $v$ and $w$.  
\item If $v\in V(G)$ appears in two subsets $\tilde{t_1},\tilde{t_2} \in T$, then $v$  
 must appear in all subsets on the (unique) path between $t_1$ and $t_2$.  
\end{enumerate} 
The width of the tree decomposition is the maximum over all nodes $t$   
in $T$ of the number of nodes in the subset $\tilde{t}$.  
The {\em tree-width} of $G$, denoted by $TW(G)$, 
 is the minimal possible width of all  
tree-decompositions of $G$.

The  {\em path width} $PW(G)$ is defined similarly, except that instead of $T$ being a tree, we constrain $T$ to be a path (i.e. a tree with all nodes of degree at most $2$). 
\end{definition}
 
It is well known that  
\begin{lemma}(Korach and Solel \cite{kor}) \label{treepath} 
Given a graph $G$ of $n$ vertices,  
$TW(G)\le PW(G)\le O(log(n))TW(G)$ where $n=|V(G)|$.  
\end{lemma}  
 
It turns out that the bubble width is tightly connected to the  
familiar notion of path-width.
 
\begin{lemma}  
Consider a graph $G$ of $n$ nodes, where each node has degree 
 bounded by an overall constant $(d)$.  
Then $\frac{1}{2}PW(G)\le BW(G)\le d\cdot PW(G)$.  
\end{lemma}  

\begin{proof}  Suppose the bubble width of $G$ is $BW(G)$ is achieved by the bubbling $b_1, b_2, \dots b_n$.  Define $\tilde{t}_i= \{ v \in G: \mbox{ an edge of $z_i(B)$ is connected to $v$ }\}$, in other words $\tilde{t}_i$ consists of all vertices that are connected to edges that cross the boundary of the bubble at the $i$th step.  It is straightforward to verify that the path of length $n-1$ which has the subset $\tilde{t}_i$ associated to its  $i$th vertex is a path decomposition; it is also clear that the path width for this decomposition is at most $2BW(G)$ and thus $\frac{1}{2}PW(G) \leq BW(G)$.

Given a path decomposition $T$ with assigned subsets $\tilde{t}_1, \tilde{t_2} \dots $ create a bubbling of the vertices of $G$ as follows:  
list,  in any order, the vertices in $\tilde{t}_1$, then list in any order those vertices in $\tilde{t}_2$ that were not in $\tilde{t}_1$, then list those vertices in $\tilde{t}_3$ that are not in $\tilde{t}_2$ in any order, etc.  Let $b_1, b_2, \dots b_n$ be this order of the vertices.  We analyze the width of this bubbling.  For each $i$, let $j_i$ be the index of the first $\tilde{t}_{j_i}$ for which $b_i \in \tilde{t}_{j_i}$.  For any edge $(a,b)$ with $a\in S_i=\{b_1, \dots b_i\}$ and $b \not\in S_i$, notice that it must be the case that  $a\in \tilde{t}_a$ for some $a \leq i_j$ and $b\in \tilde{t}_{b}$ for some $b\geq i_j$.  It follows from the conditions on path decompositions, that we must have $a\in \tilde{t}_{j_i}$.  Thus for every edge in $z_i(B)$, at least one of the vertices is contained in $\tilde{t}_{j_i}$.  It follows then that $|z_i(G)| \leq d |t_{j_i}|$ and thus $BW(G) \leq d PW(G)$. 
\end{proof}

We can combine the above two lemmas to obtain the following statement:

\begin{lemma} \label{equiv}
The three parameters, bubble-width, path-width, and tree-width are  
equal up to polylogarithmic factors.
\end{lemma}

\section{Labeled Graphs and Operator Circuits} 
\begin{definition} Given a finite graph $G$, an {\em edge labeling} $l$ of $G$ will be a map $l: E(G) \rightarrow \{0,1\}$.  If $H$ is a subgraph of $G$, then a labeling of $G$ induces a labeling of $H$, we will refer to this labeling of $H$ by $l$ as well.\end{definition}

Let $A$ be a two dimensional vector space with orthonormal basis vectors $|0 \ra$ and $|1 \ra$.  For a set of edges $E$ of some graph, we shall let $A^{\otimes E}$ be the vector space of the tensor product of $|E|$ copies of $A$, one corresponding to each element of $E$.  For a labeling $l$ of $E$, the notation $\alpha^{l(E)}$ shall mean the basis vector of $A^{\otimes E} $ corresponding to the tensoring together of the basis element $| l(e) \ra $ in the component of $A^{\otimes E}$ corresponding to the edge $e$.  Thus the set of $\alpha^{l(E)}$ as $l$ ranges over all labelings of $E$ is an orthonormal basis of $A^{\otimes E}$.

\begin{definition}  Given a finite graph $G$, for each vertex $v \in G$, a tensor associated to $v$ shall be a map $m_v$  from the set of labelings of $G_v$ to $\C$. 
\end{definition} 

The tensor $m_v$ induces many linear maps which we describe here.   Let $E=E(G_v)$ be the set of edges adjacent to $v$.  
Then $m_v$ determines a linear map  $m_v^{E, \emptyset}: 
\ A^{\otimes E}\rightarrow \C$ given by the equation 
\[m_v^{E,\emptyset}( \alpha^{l(E)} )=m_v(l), \]
 for all labelings $l$ of $G_v$.   In addition, for 
any partition of $E$ into two sets, $E=E_1 \coprod E_2$, $m_v$ 
determines a linear map $m_v^{E_1,E_2}:A^{\otimes E_1}
 \rightarrow A^{\otimes E_2}$ implicitly determined by 
the equation that for all labelings $l$ of $G_v$, 
\[ <m_v^{E_1, E_2} (\alpha^{l(E_1)}), \alpha^{l(E_2)}>= m_v(l), \]
 where $l(E_i)$ is the labeling of $E_i \subset E$ induced 
by the label $l$ of $E$.  Finally, we define the 
map $m_v^{\emptyset, E}: \C \rightarrow A^{\otimes E}$ given by  
\[ m_v^{\emptyset, A} (1)= \sum_{l \mbox{: $l$ is a label of $G_v$}}
 m_v(l) \alpha^{l(A)} . \]

\begin{definition} Given a finite graph $G$, a {\em tensor assignment m} to $G$ will be a specification of a tensor $m_v$ to  every  vertex $v$ of $G$. 
\end{definition}

\begin{definition}{\bf Tensor Circuit}  A {\em Tensor
 circuit} $T=(G,M)$ shall be any graph $G$ with a tensor 
assignment $M$.  The value of the circuit, denoted $T(G,M)$ shall 
be defined as follows: 
\[ \sum_{\mbox{ $l$ : $l$ is a label of } G} 
\ \  \prod _{v\in V(G)} m_v(l). \] 
\end{definition} 

\section{From Operator circuits to Tensor Circuits} \label{s:otot}
 
We would like to associate with an operator circuit $Q$ and an input string $x$, 
a tensor circuit $T_Q$. We shall do this in two steps, first we modify the operator circuit $Q$ to a new operator circuit $Q'$, then we associate a tensor circuit to $Q'$.  Given a linear gate $g: A^m \rightarrow A^n$ going from $m$ to $n$ qubits, we define the adjoint gate $g^*: A^n \rightarrow A^m$ that is determined by the following:  for all $x\in A^n$ and $y\in A^m$, $\la y| A^* x \ra = \la A y| x \ra $.  Given an operator circuit $Q$, define the operator circuit $Q'$ as follows:  first apply $Q$, then apply on the answer qubit the operator that projects onto $|0\ra$, and finally  apply the "adjoint" of $Q$, i.e. the circuit that is $Q$ flipped upside down with each gate $g$ replaced by the adjoint gate $g^*$.    We leave it to the reader to verify that the inner product between an output string $x$ and  $Q'$ applied to an input string $x$ is the norm squared of $Q(x)_0$ (recall $Q(x)_0$ denotes the vector that is the projection of $Q$ applied to $x$ onto the subspace that has the answer qubit $0$).  We now describe the tensor circuit $T_Q(G,M)$.  The graph $G$ shall be the graph associated with the circuit $Q'$.  The tensor assignment $M$ is as follows:

\begin{itemize}
\item For a vertex $v$ of $G$ corresponding to a linear gate $g: A^n \rightarrow A^m$  of $Q'$,  let $E_1$ (respectively $E_2$) be the edges in $G$ corresponding to the $n$  input qubits (respectively $m$ output qubits) that meet at $v$.   We assign to $v$ the tensor $m_v$ for which the associated linear map $m_v^{E_1, E_2}$ is the linear gate $g$.
\item For the vertex $v$ of $G$ corresponding to the gate that projects onto $|0\ra$ in the answer qubit (which has degree 2) we define the tensor associated to $v$ by 
\[m_v(|0\ra |0\ra)=m_v(|1\ra|1\ra)=1, \ m_v(|0\ra |1\ra)= m_v(|1\ra |0\ra)=0. \]
\item For the vertices $v$ corresponding to the $i^{\mbox{th}}$ input or output qubit of $Q'$ (which have degree 1), we define the tensor associated to $v$ by $m_v(|x_i\ra)=1$ and $m_v(|x_i \oplus 1\ra)=0$.
\end{itemize}

With this construction we have the following connection between the operator circuit $Q$ and the tensor circuit $T_Q(G,M)$:

\begin{lemma} \label{cl:quantumvalue}  
The value of the tensor circuit $T_Q(G,M)$ defined above  is the norm squared of $Q(x)_0$.
\end{lemma}
\begin{proof}  It is straightforward to verify that the value of the tensor circuit is the inner product of $x$ with $Q'$ applied to $x$.  The result then follows from the observation made earlier that this latter inner product is equal to the norm squared of $Q(x)_0$.
\end{proof}

\section{Efficient Simulations of Operator Circuits of logarithmic 
Bubble-Width} \label{s:3}

We want to prove Theorem \ref{thm:main}. 
We start by moving from an operator circuit $Q$ and an input vector $x$ to its tensor circuit $T_Q(G,M)$ as in the previous section. We highlight the connection between the bubble width of $G$ and that of the graph associated with $Q$: 
\begin{lemma} \label{l:octotcbubble}
Given $Q$ and $G$ as above, $BW(G) \leq 2 BW(Q) +1$.
\end{lemma} 
\begin{proof}
\znote{Dorit. . . I leave it to you to say something here}
\end{proof}

The following theorem, when combined with lemmas \ref{l:octotcbubble} and \ref{cl:quantumvalue}, implies
Theorem \ref{thm:main}:  
\begin{theorem} 
Given a tensor circuit $T(G,M)$,
its value can be computed classically
in time polynomial in $|V(G)|2^{BW(G)}$.  In particular 
if $BW(G)$ is logarithmic in $|V(G)|$ then the time is 
 polynomial in the size of the graph.
\end{theorem} 
\begin{proof} 
We will produce vectors $\psi_i$, $1\leq i \leq |V(G)|$ with $\psi_i 
\in A^{\otimes z_i(B)}$.   The last vector, $\psi_n$ 
(a scalar since $z_n(B)$ is empty) is the value of the circuit. 
 The map from $\psi_i$ to $\psi_{i+1}$ will be a linear map.  
Our result will then follow.

The main idea here is the following. 
A tensor assigned to a vertex induces many linear maps; 
we  choose the linear maps that minimize the number of computational steps. 
The choice will be determined by the best bubbling.   Let $b_1,\ldots,b_n$ be the bubbling of $G$ which achieves the 
bubble width of $G$.  Now let $\psi_1 = 
m_v^{\emptyset, z_1(B)} (1) \in A^{\otimes z_1(B) }$, 
(note that $z_1(B)$ consists of all the adjacent edges of $v_1$).  
Given $\psi_{i-1} \in A^{\otimes z_{i-1}(B)}$ we show how to 
compute $\psi_{i}$.   Split the incident edges of $v_i$ into 
two groups $E_1$ and $E_2$ where $E_1$ is the set of edges that 
are in $z_{i-1}(B)$  and $E_2$ is the set of edges in $z_i(B)$. 
It follows that $E_1 \coprod E_2$ is the set of all 
edges incident to $v_i$ and  $z_{i-1}(B) -E_1= z_{i}(B) -E_2$.  We 
now set $\psi_{i}= \overline{m_v^{E_1, E_2}} (\psi_{i-1})$, 
where $\overline{m_v^{E_1, E_2}} : A^{\otimes z_{i-1}(B)} 
\rightarrow A^{\otimes z_i(B)}$ is the linear map that is 
the identity on $A^{\otimes z_{i-1}(B) - E_1}= A^{\otimes z_i(B)-E_2}$ 
tensor with the linear map $m_v^{E_1,E_2}: A^{\otimes E_1} 
\rightarrow A^{\otimes E_2}$. 
 
We leave it to the reader to verify that with these definitions, $\psi_n$
ends up being the value of the tensor circuit $T(G,M)$. 
 
The complexity of this algorithm is the sum of the complexities of the application of the linear maps that take $\psi_i$ to $\psi_{i+1}$.  
There are $|V(G)|$ such linear maps and the largest 
vector space encountered is the tensor product of $BW(G)$ 
copies of $A$ and is thus of dimension $2^{BW(G)}$. 
It follows that the complexity is polynomial in $|V(G)|2^{BW(G)}$. 
\end{proof}

\section{Fourier Transform Circuit of logarithmic Bubble width}  
We shall modify the construction by Cleve and Watrous \cite{cw}  
of the log-depth  
 quantum circuits for Fourier transform to produce a circuit of poly-logarithmic bubble width. 
The modification takes advantage of the fact with the more general linear operator circuits,  bits can be  
{\it erased} easily.  
In other words - the transformation  
\begin{equation}  
|0\ra,|1\ra\longmapsto 1 
\end{equation} 
where $1$ is simply a scalar, is a valid transformation.  
 
We are interested in constructing an operator circuit that performs an approximation of  the quantum Fourier transform.  We begin with notation consistent with \cite{cw}.   By $|x \ra $ we shall mean the basis state $|x \ra= |x_{n-1} \ra | x_{n-2} \ra \cdots |x_0 \ra$.   We define  $|\mu _{\theta} \ra= \frac{1}{\sqrt{2}}( |0\ra + e^{2\pi i \theta} |1\ra)$.  Then the quantum Fourier transform is the linear extension of the  map
\[ |x\ra \rightarrow |\psi_x \ra= |\mu_{0.x_0}\ra|\mu_{0.x_1x_0 }\ra \dots | \mu_{0.x_{n-1}x_{n-1} \dots x_0} \ra .\]

We remark that as in \cite{cw}, the state 
\[ |\tilde{\psi}_x\ra= |\mu_{0.x_0} \ra | \cdots |\mu_{0.x_{k-1}}\cdots x_0\ra |\mu_{0. x_k \cdots x_1}\ra |\mu _{0.x_{k+1}} \cdots x_2 \ra \cdots  |\mu_{0.x_{n-1} \cdots x_{n-k} }\ra , \] where we replace each $\mu_{\theta}$ by the approximation of $\theta$ by the first $k$ digits after the decimal point, is a good approximation for $| \psi _x \ra$ when $k=2log(n/\epsilon)+O(1)$.  Our construction will be of a circuit that applies the linear extension of the map $|x\ra \rightarrow 
 |\tilde{\psi}_x \ra$.
 
 Our circuit will be composed of the product of three circuits applied sequencially:
 \begin{enumerate}
 \item The linear extension of the map defined by 
 \[|x \ra \rightarrow |\alpha_x \ra= |x_n^k\ra|0^k\ra|x^k_{n-1}\ra|0^k\ra \cdots |x^k_0\ra |0^k\ra.\]
\item The linear extension of the map defined by
\[ |\alpha _x \ra \rightarrow |\beta_x\ra= |x_n^k\ra |0^{k-1}\ra|\mu_{0.x_0} \ra |x^k_{n-1}\ra |0^{k-1}\ra|\mu_{0.x_1 x_0} \ra  \cdots |x^k_n\ra|0^{k-1}\ra |\mu_{0.x_n \dots x_{n-k} }\ra.\]
\item The linear extension of the map defined by
\[ |\alpha_x\ra  \rightarrow |\tilde{\psi}_x\ra. \]
\end{enumerate}

The first map is straightforward.  We denote the map that makes one copy of a single qbit, i.e. the linear extension of the map defined by $|0\ra \rightarrow |0\ra |0\ra$, $|1\ra \rightarrow |1\ra|1\ra$ by the picture
\[ \hpic{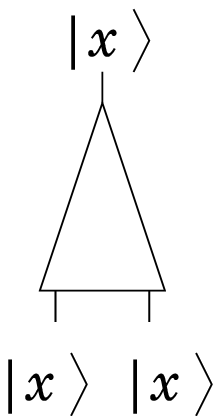} {1in} . \]
Then we can create $k$ copies of each bit  with a $\log k$ depth circuit  by using  
$O(k)$ of these maps, as in the following picture:   
 
\[ \hpic{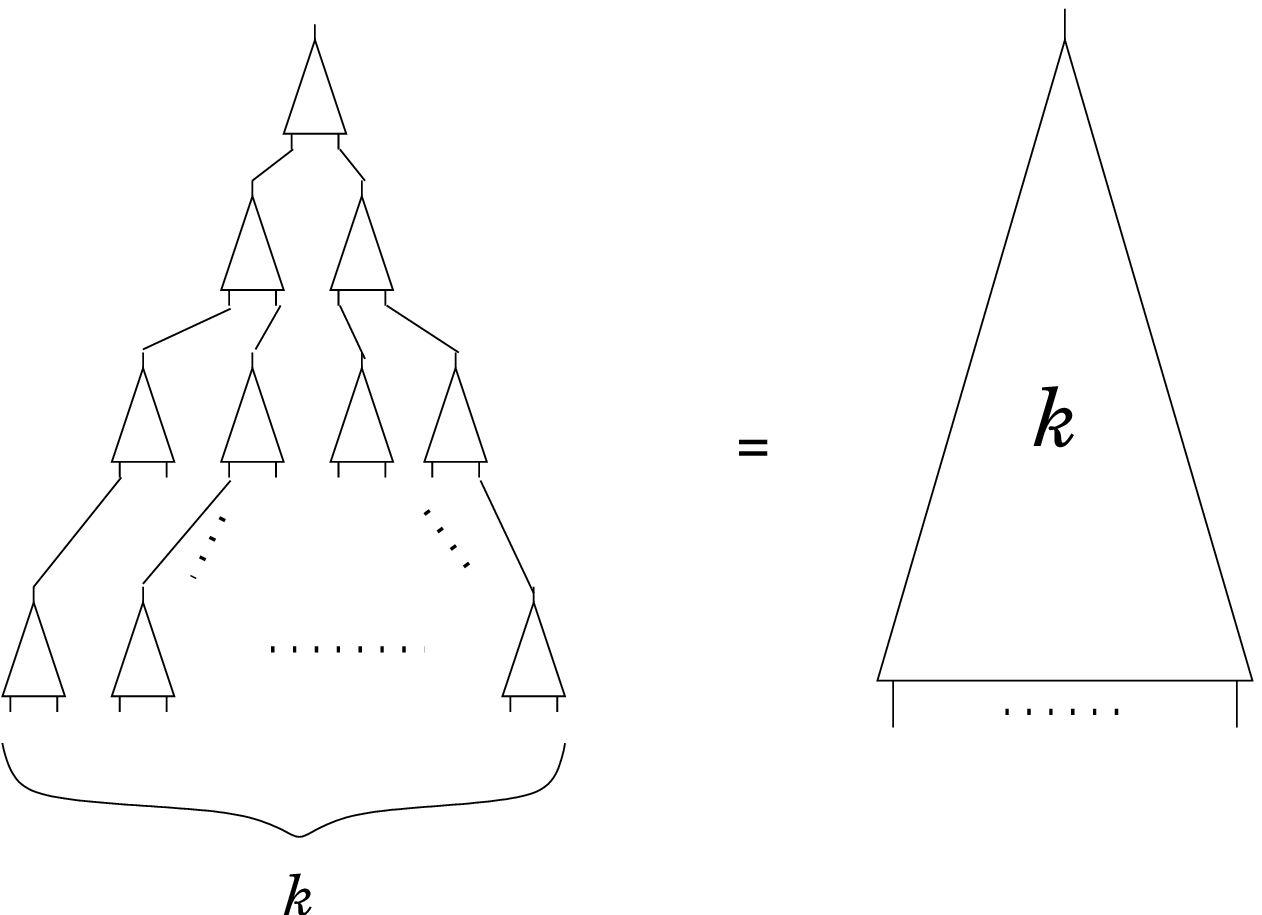} {2 in} \] 
 
Now, we insert an extra bunch of $k$ qubits  
in the state  
$|0\ra$ to the right of each bunch of copied qubits,  
 using the linear  
operator $1\longmapsto |0\ra$.

The third map is also straightforward since we are using linear circuits and we do not require unitarity of the gates.  Notice that $|\tilde{\psi}_x \ra$ can be gotten by eliminating all bits  except those in the $2k$th, $4k$th, $6k$th etc location.   
Unlike in the unitary case, where a lot of effort  
was put into getting rid of the remaining so called computational bits,   
here we independently at each location apply the simple transformation  
which takes all those bits to the scalar $1$.

The more involved component is the second circuit.  Following \cite{cw}, the circuit below is the linear extension of the map:
\[ |x\ra |0^k\ra \rightarrow |x \ra|0^{k-1}\ra |\mu_{0.x_jx_{j-1} \dots x_{j-k+1}} \ra, \]

\noindent which can be implemented according to the following diagram:

 \[ \hpic{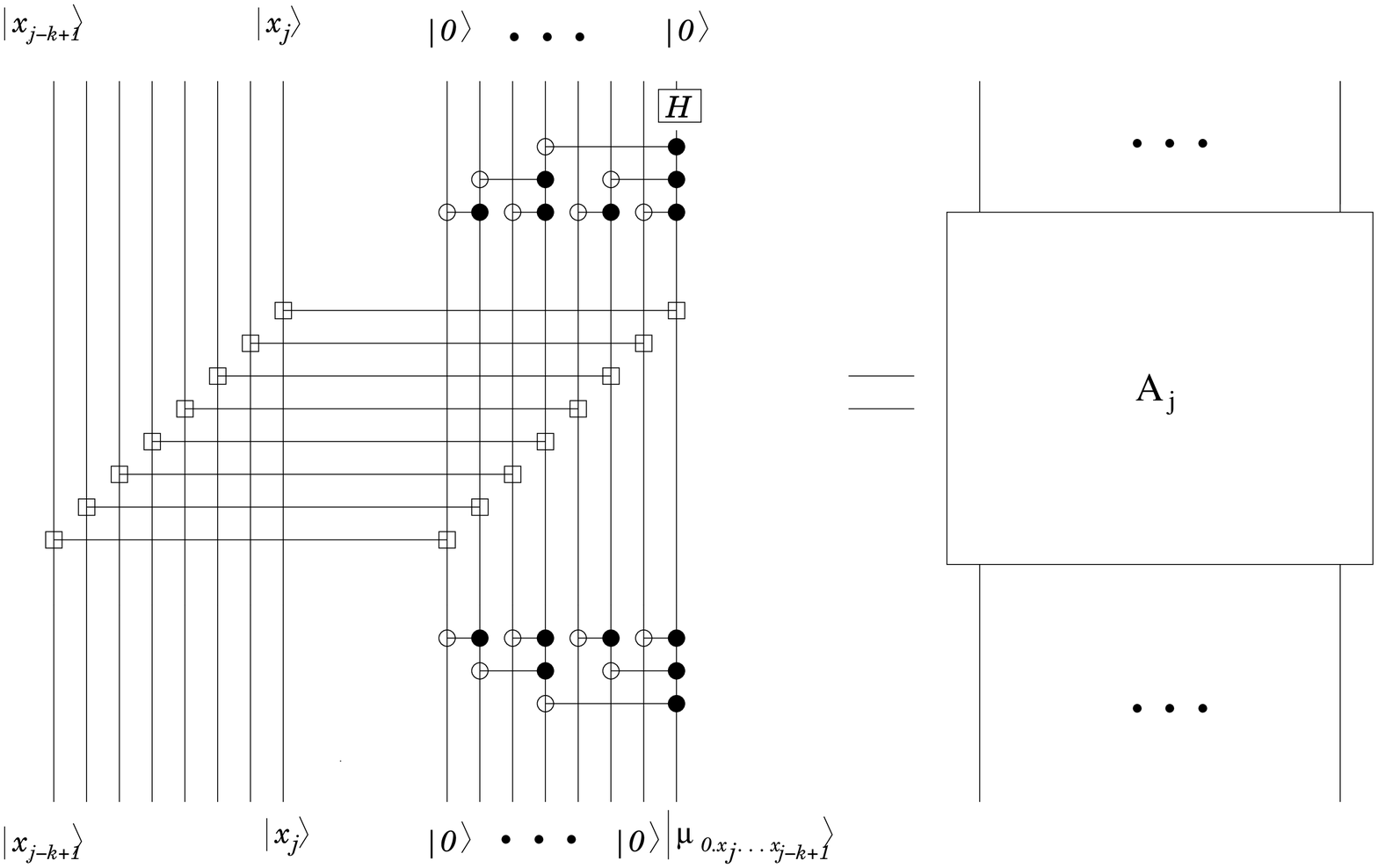} {3 in} \]

\noindent where the "H" gate is the Hadamard gate and the gates with one open and one closed circle are C-NOT gates.  The gates with two open squares, though depicted as identical to each other,  are different controlled-phase shift gates which we now describe.  Define the controlled-phase shift map $c-P(\theta)$ to be the map defined by $|x \ra |y \ra \rightarrow e^{2\pi i \theta x \dot y}|x\ra |y\ra$.  Then the two open square gate in the above diagram that acts on $|x_l\ra$ is $c-P(2^{l-j+1})$.

Thus to implement the second map we apply in parallel the gates $A_j$, $0\leq j \leq n-1$ to the state $|\alpha _x\ra$ in the following way:  $A_j$ acts on the strands of $|\alpha_x\ra$ corresponding to the  $k+1$th copy of each $|x_i\ra$ it needs as well as those corresponding to the $k+1$ th {\it block } of $|0^k\ra$.  Thus the $A_j$ act on disjoint sets of strands (for different $j$) and therefore they can be applied in parallel.  We note that each $A_j$ has "width" no bigger than $2k^2$, i.e. the distance between any two strands that $A_j$ acts on is no more than $2k^2$.  It should be clear that the application, in parallel in this way, of the gates $A_j$, $0\leq j \leq n-1$ implements the map $|\alpha_x \ra \rightarrow |\beta_x \ra$.

We have completed the description of the circuit that implements the approximation of the Fourier transform.  It is left to upper bound the bubble width of the above operator circuit.  
To do this we describe a certain bubbling which will provide an upper bound  
on the bubble width.  
The bubbling is very simple: we bubble from left to right.  The precise order does not matter as long as the bubbling swallows gates above and below things it has already swallowed before swallowing too many things farther to the right.  The resulting width for this bubbling is no more than quadratic 
in $k$ (and thus by choice of $k$ poly-logarithmic).  
The reason for this relies on two features of the 
circuit: a) the circuit has depth linear in $k$, and b) the 
"width" of any gate encountered is no more than quadratic in $k$.

%
 
This completes the proof of Theorem \ref{thm:fft}. 

\section{Remarks on why the simulation fails for Shor's algorithm}  
It is natural to ask whether these techniques can be extended 
to provide an efficient classical simulation of Shor's algorithm. 
All our attempts to do so have failed, and it seems that there is an 
inherent difficulty here. 
The reason is that the modular exponentiation part in the algorithm requires  
multiplication, and to the best of our knowledge, the bubble width of  
multiplication circuits is close to linear. One might hope to
try and avoid this problem by using simpler operations that would 
suffice for factoring. 
However, all our attempts to do so encountered yet another problem which seems 
difficult to handle:  
the bubble width is not additive. One can connect circuits  
of very small bubble width, to get a very large bubble width. Hence, 
 not only that the different parts of the factoring 
circuit need to have small bubble width, 
but their connections need to be designed in such a way that the bubble width 
of the entire circuit is still small.

\section{Epilogue: The proof of Theorem \ref{thm:main}
 using the Jones polynomial technique} \label{s:4} 
 
Here   
 we sketch an alternative, short proof of Theorem \ref{thm:main} in the case when the operators involved in the circuit are unitary.  
We assume familiarity with the  
notions of the Jones polynomial, braids, and the statements of the  
recent results in quantum computation regarding these notions \cite{ajl,arad}. 
More background can be found in \cite{ajl} and \cite{arad}.  
The proof is achieved by combining the  
quantum universality of the Jones polynomial \cite{freedman,arad} 
with the well known fact that the Jones polynomial of a braid 
can be calculated in time at most exponential in the tree-width of  
the graph underlying the braid \cite{noble,and}.

{\it Proof:}
Given a quantum circuit $Q$ on $n$ qubits and with $s$ gates,  
whose bubble-width is poly-logarithmic, we perform the following steps:

\begin{enumerate}
\item \label{step1} 
We  create a quantum circuit  
$Q'$, of $n'$ qubits and $s'$ gates, such that: a) $n',s'$ are at most polynomial  
in $n,s$, b) the probability that $Q$ outputs $0$ is equal to  
$\la 0^n|Q'|0^n\ra$. 
\item \label{step2}We create a braid $b$  
whose Jones  
polynomial at a particular root of unity is inverse-polynomially close to $\la 0^n|Q|0^n\ra$.   The graph corresponding to $b$ will have poly-logarithmic bubble width.
\item \label{step3} We classically evaluate the Jones polynomial at the particular root of unity in quasi-polynomial time.
\end{enumerate}

Step \ref{step1} is a standard construction in quantum 
computation-see \cite{nielsen} or if you are really desperate,  
pages $9-10$ in \cite{arad} (we note it is very similar to the construction from $Q$ to $Q'$ in section \ref{s:otot}).  It is simple to 
verify that the bubble width of $Q$ is no more than one more than twice the 
bubble width of $Q'$ (again, this is the same result as lemma \ref{l:octotcbubble}).  Step \ref{step2} follows from the 
results of \cite{freedman,arad}.  Specifically, the 
braid $b$ has $4n$ strands, and each gate in the original circuit is replaced by poly-logarithmically  
many crossings in the braid $b$, on the $4$ or $8$ 
strands corresponding to the qubit or qubits involved in the gate.  
 It is straightforward to see that the bubble-width of the 
underlying graph of the braid (the underlying graph is the graph obtained by replacing every crossing by a vertex) remains poly-logarithmic.  Consequently, Lemma \ref{equiv} implies that the tree-width of the underlying graph of this braid is poly-logarithmic as well.  Step \ref{step3} follows from the
 known result  \cite{noble,and}
which states that the Jones polynomial at any point, of  
a braid whose underlying graph has poly-logarithmic tree-width,  
can be calculated in time which is quasi-polynomial.

\appendix 
   
  \end{document}